\documentclass[review]{elsarticle}

\usepackage{multirow}

\journal{Arxiv}

\begin{document}

\begin{frontmatter}

\title{Sensitivity and Specificity Evaluation of Deep Learning Models for Detection of Pneumoperitoneum on Chest Radiographs}

\author[1]{Manu Goyal\corref{cor1}}
\cortext[cor1]{Corresponding author}
\ead{manu.goyal@dartmouth.edu}
\author[2]{Judith Austin-Strohbehn}
\author[2]{Sean J. Sun}
\author[2]{Karen Rodriguez}
\author[2]{Jessica M. Sin}
\author[2]{Yvonne Y. Cheung}
\author[3]{Saeed Hassanpour}

\address[1]{Department of Biomedical Data Science, Dartmouth College, Hanover, NH, USA.}
\address[2]{Department of Radiology, Dartmouth-Hitchcock Medical Center, Lebanon, NH, USA.}
\address[3]{Departments of Biomedical Data Science, Computer Science, and Epidemiology, Dartmouth College, Hanover, NH, USA.}

\begin{abstract}
\textbf{Background:} Deep learning has great potential to assist with detecting and triaging critical findings such as pneumoperitoneum on medical images. To be clinically useful, the performance of this technology still needs to be validated for generalizability across different types of imaging systems.

\noindent\textbf{Purpose:} To evaluate the performance of deep learning to detect pneumoperitoneum in chest radiographs and to conduct sensitivity and specificity analysis of common deep learning architectures across different types of imaging systems at various institutions.

\noindent\textbf{Materials and Methods:} This retrospective study included 1,287 chest X-ray images of patients who underwent initial chest radiography at 13 different hospitals between 2011 and 2019. State-of-the-art deep learning models were trained on a subset of this dataset, and the automated classification performance was evaluated on the rest of the dataset by measuring the AUC, sensitivity, and specificity. Furthermore, the generalizability of these deep learning models was assessed by stratifying the test dataset according to the type of imaging systems utilized. 

\noindent\textbf{Results:} All deep learning models performed well for identifying radiographs with pneumoperitoneum, while DenseNet161 achieved the highest AUC of 95.7\%, Specificity of 89.9\%, and Sensitivity of 91.6\%. The DenseNet161 model was able to accurately classify radiographs from different imaging systems (Accuracy: 90.8\%), while it was trained on images captured from a specific imaging system from a single institution. This result suggests the generalizability of our model for learning salient features in chest X-ray images to detect pneumoperitoneum, independent of the imaging system.

\noindent\textbf{Conclusion:} DenseNet161 achieved high performance for the detection of pneumoperitoneum, a result that holds across different types of imaging systems. If verified in clinical settings, this model could assist practitioners with the diagnosis and management of patients with this urgent condition.

\end{abstract}


\end{frontmatter}


\section{Introduction}
In recent years, advances in deep learning have presented new opportunities to assist and improve clinical diagnosis involving different medical imaging modalities such as magnetic resonance imaging (MRI), X-ray, Ultrasound, computed tomography (CT), and positron emission tomography (PET) \cite{ahmad2018semantic, yan2018weakly,  yap2020breast, tomita2018deep, liu2018deep}. Chest X-rays (CXR) are commonly used as an important imaging tool to screen patients for a number of diseases. 

In recent studies, deep learning has provided end-to-end proof of concept for models achieving radiologist-level performance in the detection of different clinical findings on CXRs \cite{rajpurkar2017chexnet, wang2017chestx}. In doing so, deep learning has helped physicians to prioritize urgent medical cases and focus their attention during the course of clinical diagnosis.

Pneumoperitoneum is a critical clinical finding that requires immediate surgical attention \cite{stapakis1992diagnosis, woodring1995detection}. Although abdominal radiographs and CT scans are standard modalities for the detection of pneumoperitoneum, CXRs are often an initial exam that is ordered in the emergency room setting. Therefore, pneumoperitoneum is often detected on initial CXRs, before additional imaging, such as CT exams, are ordered. Free air in the abdomen is most visible on CXRs of patients in the standing position. Because gas ascends to the highest point in the abdomen, free air accumulates beneath the domes of the diaphragm in the standing or upright position. Therefore, CXR is one of the most sensitive modalities to detect pneumoperitoneum \cite{chen2002ultrasonography}. Solis et al. showed that performing abdominal CT exams can delay surgery, without providing any measurable benefit over a CXR for the diagnosis of pneumoperitoneum \cite{solis2014free}. 

Despite the recent success of deep learning models in detecting disease on CXRs, it has been found that these models can be highly sensitive to the types of systems used for the training dataset. For instance, Marcus et al. \cite{marcus_little_2019} argued a deep learning model trained on standard CXR images captured by a particular imaging system in a fixed location may not perform as well on portable CXR images. This is because the trained deep learning model has to deal with variabilities in patterns and characteristics found in CXR images across different imaging systems, rather than variability and differences in chest anatomy and morphology intrinsic to the disease itself. In this study, we developed state-of-the-art deep learning models to detect pneumoperitoneum on CXR images and evaluated the sensitivity and specificity of these models on a diverse dataset assembled from different types of X-ray imaging systems from various hospitals to demonstrate the generalizability of our approach.

In a hemodynamically stable patient with acute, severe, or generalized abdominal pain, multidetector CT scan is the preferred imaging test and it provides invaluable diagnostic information in the diagnostic workup of such patients \cite{paolantonio2016multidetector}. However, in unstable patients, the emergency room staff, or clinical house staff, need timely information. In these cases, usually, clinicians first obtain an upright CXR to exclude a critical finding of free air when gastrointestinal perforation is suspected. Once the patient is stabilized, a CT scan is also usually obtained, but only after an initial plain radiograph of an upright CXR is performed to quickly look for pneumoperitoneum.  In this study, we focused on the CXR exam, since it is one of the most common initial radiographs to be performed in hospital settings, including emergency rooms and inpatient and outpatient settings for patients with acute, severe, and generalized abdominal pain. The purpose of the deep learning tool is to assist radiologist readers with prioritizing interpretations of the most urgent exams and help them to reach a prompt, correct diagnosis.

In this study, CXRs were performed with fixed imaging, which is performed in the X-ray department, as well as with portable imaging, which is performed with mobile X-ray units. The portable CXR is useful for diagnosis and monitoring patients at their bedside in the emergency room, and in the intensive care unit or inpatient setting, which are often utilized when a patient cannot be transferred to the hospital radiology department. Despite the advantages of portable CXR, the image quality of a portable bedside CXR can be limited, and the image interpretation and appropriate clinical action can be affected. Therefore, this study aims to determine if there are differences between portable CXRs and fixed CXRs performed in a radiology department in terms of the detection of pneumoperitoneum. Additionally, given the image variabilities due to differences in scanners from different manufacturers, such as GE, Philips, and Siemens, that are used to obtain CXRs, this study aims to determine if there was a difference in detection of pneumoperitoneum on CXRs from different manufacturers of X-ray machines.

\begin{figure*}
	\centering
	\includegraphics[scale=.75]{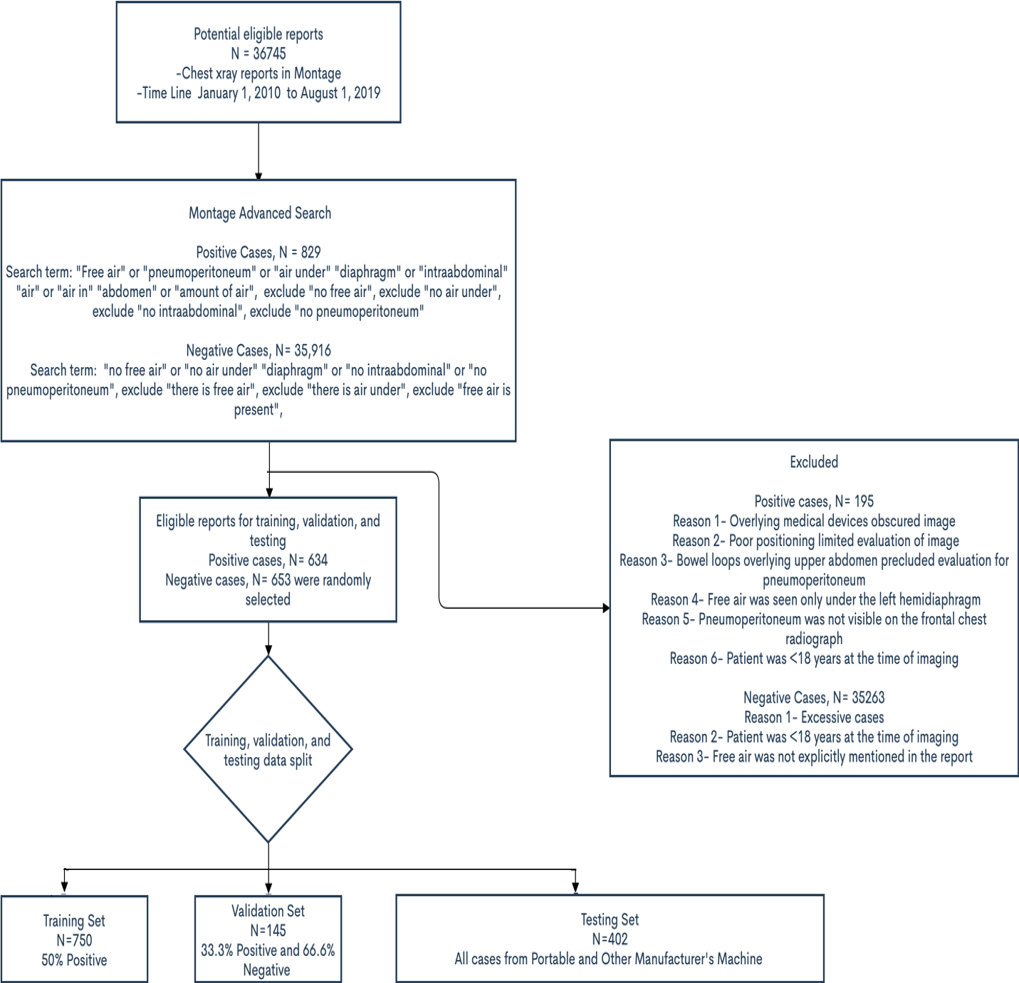}
	\caption{Details of inclusion and exclusion criteria of this study.}
	\label{fig:InEx}
\end{figure*}

\begin{figure}[!t]
	\centering
	\includegraphics[scale=0.7]{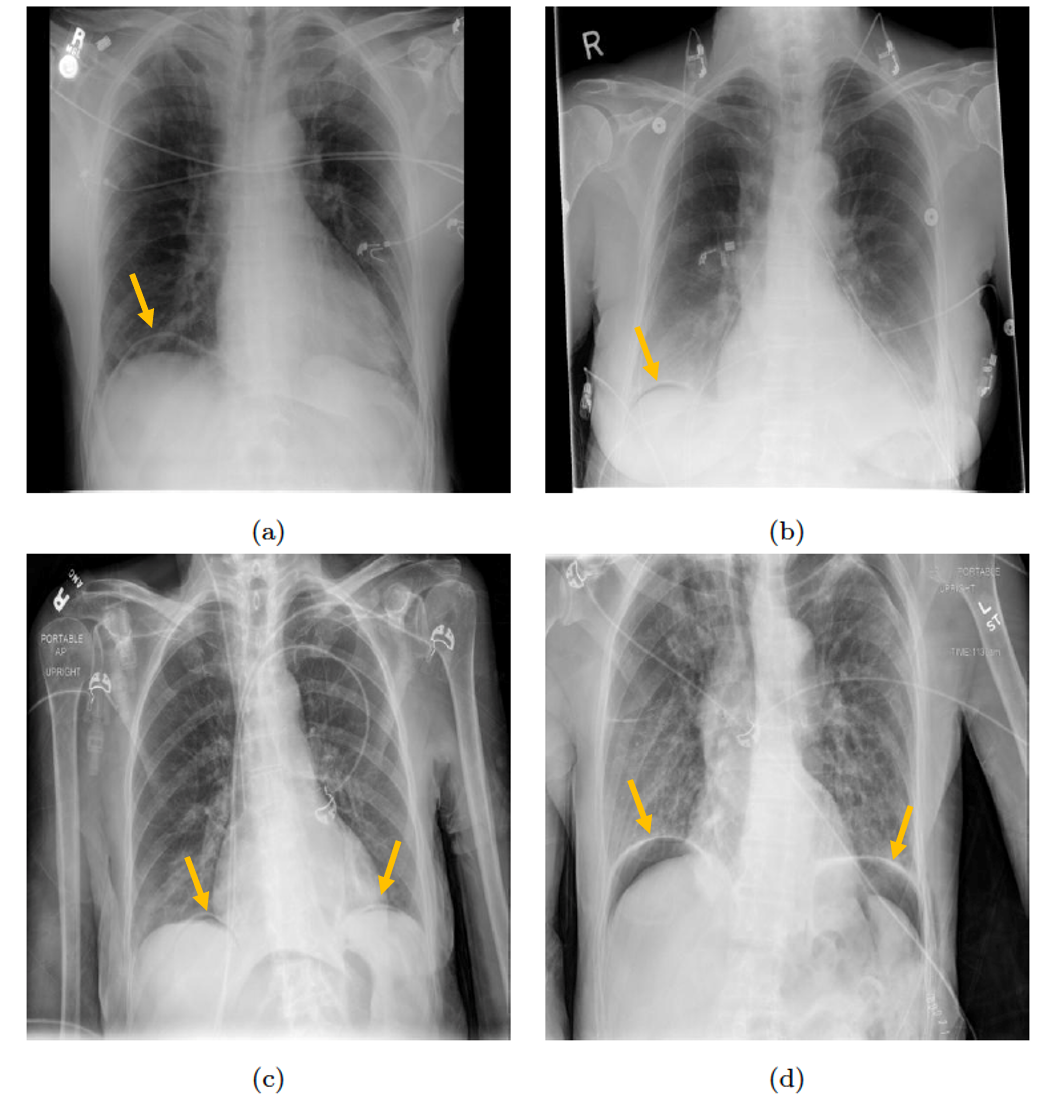}
	\caption{Four examples of pneumoperitoneum positive cases in chest X-ray images in our dataset. Where yellow arrow indicates the presence of Pneumoperitoneum.}
	\label{fig:xrayd}
\end{figure} 

\begin{figure}
	\centering
	\includegraphics[scale=0.52]{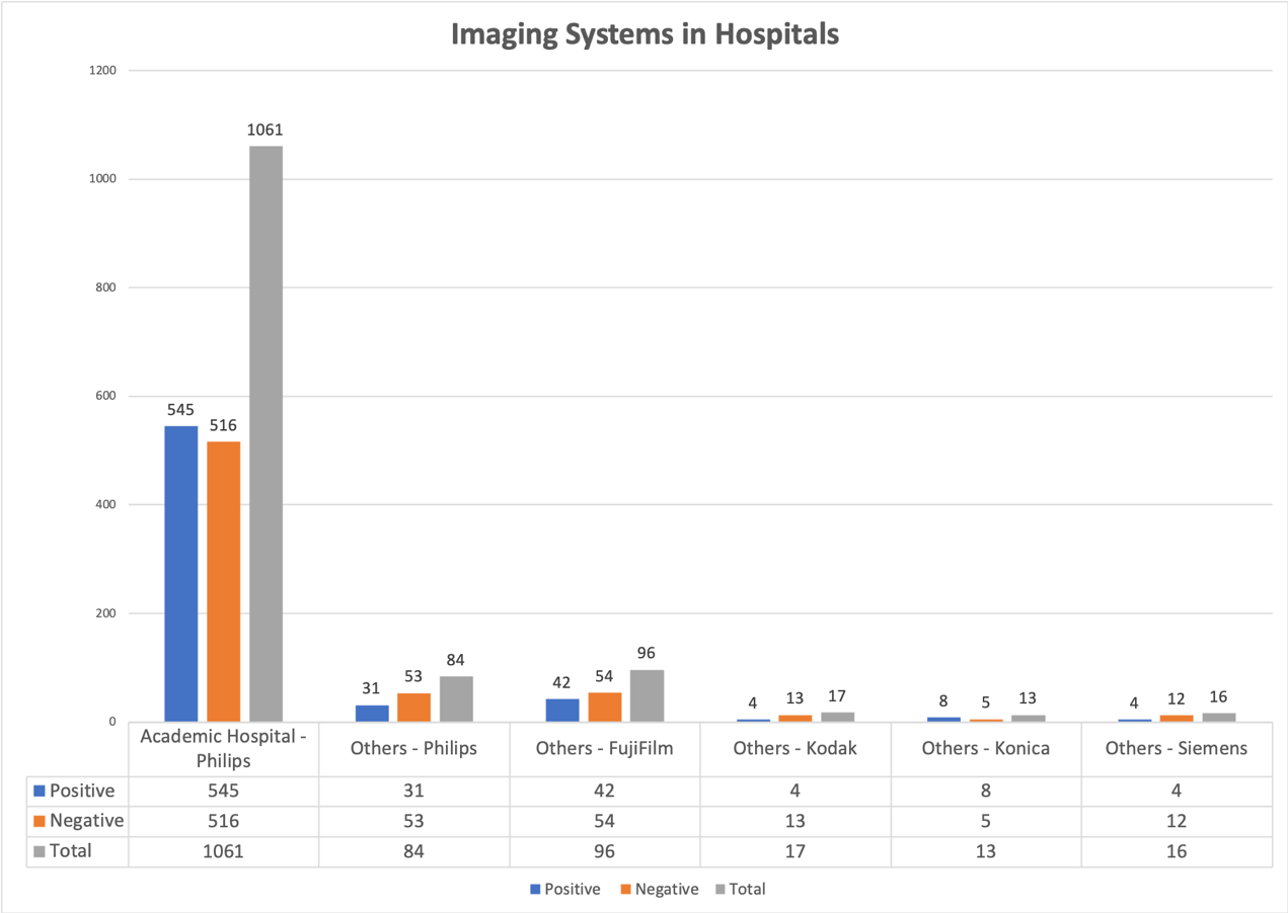}
	\caption{Number of CXRs stratified in our dataset by their corresponding imaging system manufacturer.}
	\label{fig:ImgSys}
\end{figure}

\section{Pneumoperitoneum Dataset}
The pneumoperitoneum dataset consisted of 1,287 CXR images (from 1,124 patients) and was collected using Montage (Montage Healthcare Solutions, Philadelphia, PA) search functionality from the database of a tertiary academic hospital and several community hospitals serving rural populations between March 2011 and September 2019. The inclusion and exclusion criteria for this study is demonstrated in Fig. \ref{fig:InEx}. This dataset is nearly balanced with 634 pneumoperitoneum positive cases and 673 pneumoperitoneum negative cases. The pneumoperitoneum negative cases consist of both normal and other conditions (such as pneumothorax, pneumonia, atelectasis, etc.). A brief description of this dataset is presented in Table \ref{my-label2}. All CXR images in this dataset were retrieved in DICOM format. The resolution of CXR images in our dataset ranges from 1728x1645 pixels to 4280x3520 pixels. A few examples of positive pneumoperitoneum CXR images are shown in Fig. \ref{fig:xrayd}. All CXR images from the academic hospital were taken with Philips imaging systems, whereas CXR images from other community hospitals were taken with imaging systems from various manufacturers (Philips, Fujifilm, Siemens, Kodak, Konica Minolta). Further details of the total number of CXR images specified by the imaging system manufacturer are shown in Fig. \ref{fig:ImgSys}.

\begin{table*}[]
	\centering
	\addtolength{\tabcolsep}{2pt}
	\renewcommand{\arraystretch}{1.5}
	\caption{Characteristics of our dataset stratified by pneumoperitoneum positive and negative cases. Where CXRs is Chest X-rays, SD is Standard Deviation, AP is Anteroposterior, PA is Posteroanterior, and AH is Academic Hospital}
	\label{my-label2}
	\scalebox{1}
	{
	\begin{tabular}{llll}\hline
		Characteristic                                                                 & Full Dataset   & Positive Cases  & Negative        \\\hline \hline
		No. of CXRs                                                                    & 1,287          & 634             & 653             \\\hline
		\multirow{2}{*}{Sex}                                                           & Male - 697     & Male - 344      & Male -353       \\
		& Female -590    & Female - 290    & Female -303     \\\hline
		\multirow{4}{*}{Age (SD)}                                                      & Male           & Male            & Male            \\
		& 61.44+/-17.62 & 61.06+/-17.44 & 61.82+/-17.79 \\
		& Female         & Female          & Female          \\
		& 62.03+/-17.96  & 65.09+/-16.09   & 58.98+/-19.89   \\\hline
		\multirow{2}{*}{\begin{tabular}[c]{@{}l@{}}Technique \\ (AP/ PA)\end{tabular}} & AP - 554       & AP - 251        & AP - 304        \\
		& PA - 733       & PA - 383        & PA - 429        \\\hline
		Imagine System                                                                 & Fixed - 969    & Fixed - 451     & Fixed - 518     \\
		Type                                                                           & Portable - 318 & Portable - 183  & Portable - 135  \\\hline
		\multirow{2}{*}{Hospital}                                                      & AH - 1061      & AH - 545        & AH - 516        \\
		& Others - 226   & Others - 89     & Others - 137    \\\hline
		\multirow{2}{*}{Manfacturer}                                                   & Philips - 1145 & Philips - 576   & Philips -569    \\
		& Others - 142   & Others - 58     & Others - 84   \\\hline 
	\end{tabular}}
	\end{table*}

Although several images from the same patient were included, they were not identical in terms of positioning and appearance. Since our goal in this study is to detect pneumoperitoneum per exam, multiple images for a patient do not alter the findings. Furthermore, we kept all the images from same patient in one partition (training, validation, testing). This dataset is nearly balanced, with 634 pneumoperitoneum positive cases and 673 pneumoperitoneum negative cases. For pneumoperitoneum positive cases, there is 182 post-operative pneumoperitoneum included for this study. 

\subsection{Expert Annotations}
We needed high-quality expert annotations indicating ground truth pneumoperitoneum labels (i.e., positive or negative) for each CXR image in our dataset to develop and evaluate our model. The expert annotations in our study were generated by four radiologists from the main academic hospital campus. To produce the ground truth labels, the CXR images were equally divided among two radiologists for annotation. Then, the other two radiologists independently reviewed all the ground truth labels generated by the previous radiologists for accuracy. Any disagreements among annotators were resolved by further review and discussion among all radiologists. 
We tested the consistency of expert annotation between two radiologists on 177 randomly selected test cases consisting of 45 positive and 132 negative cases. Out of 177 tested cases, there was one on which the radiologists disagreed about the presence of pneumoperitoneum. That case was negative for pneumoperitoneum, and the disagreement was resolved after further discussion among the radiologists. Radiologist 1 has over 30 years of general radiology experience; Radiologist 2 has over 7 years of general radiology experience; Radiologist 3 has 20 years of abdominal imaging experience; and Radiologist 4 is a 4th-year radiology resident.

\section{Methodology}
This section describes our proposed technique for the recognition of pneumoperitoneum in chest radiographs. The preparation of dataset, deep learning approaches used for binary classification of pneumoperitoneum are detailed in this section. 

\begin{figure}
	\centering
	\includegraphics[scale=.35]{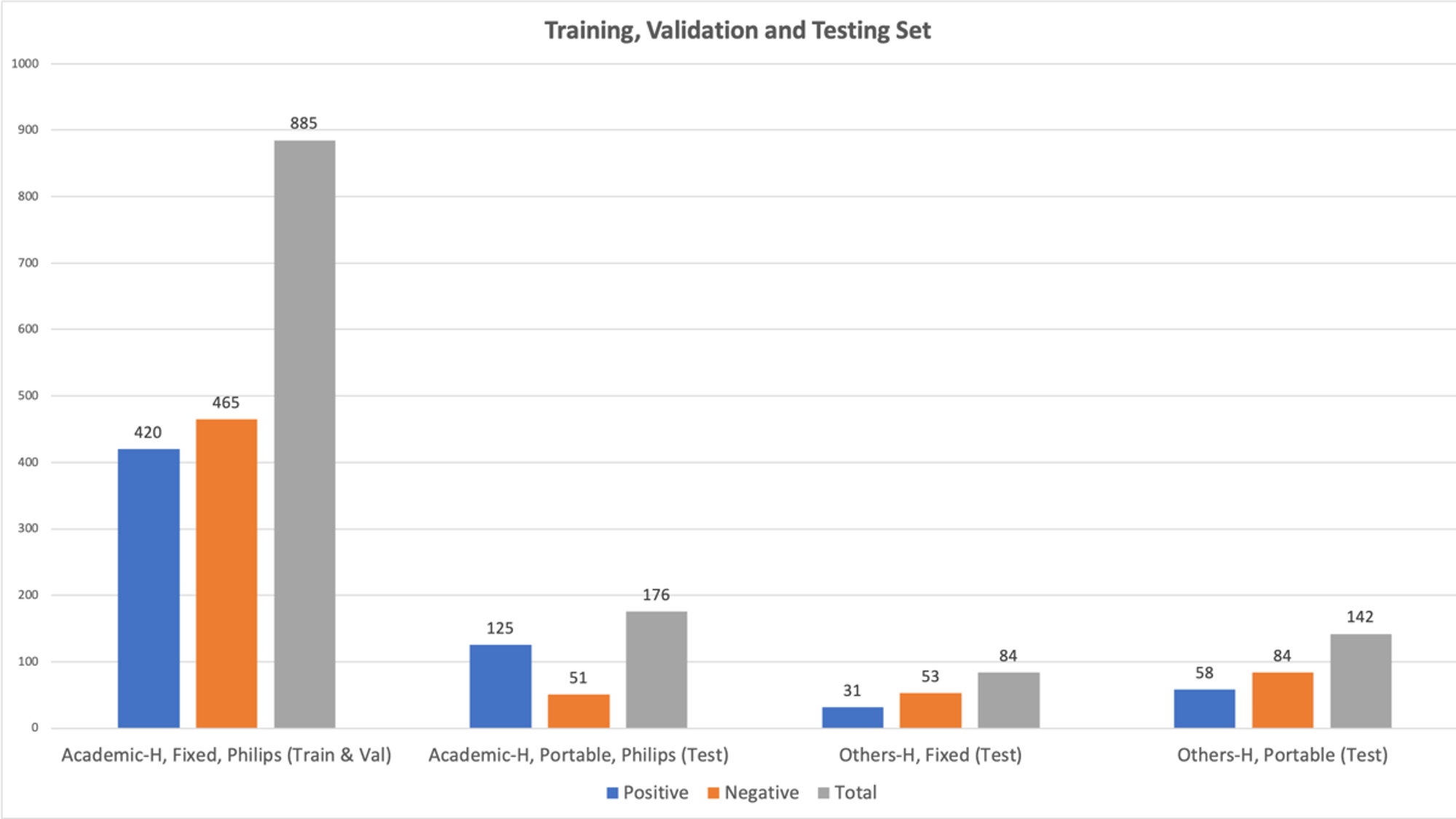}
	\caption{Number of CXRs stratified in our dataset by their corresponding imaging system manufacturer.}
	\label{fig:TrVTe}
\end{figure}

\subsection{Training, Validation and Test Dataset Split}
Our study has two objectives: 1) to train and evaluate the performance of common deep learning architectures on our CXR image dataset for classification of pneumoperitoneum status, and 2) to analyse the sensitivity and specificity of these models based on different characteristics of the radiographs. Therefore, as shown in Table \ref{my-label2}, we partitioned this dataset into training, validation, and test datasets. For the training and validation datasets, we only used the CXR images with the most common characteristics in the dataset, i.e., images taken by fixed X-ray machines at the main academic hospital (420 positive cases and 465 negative cases; Fig. \ref{fig:TrVTe}). The training dataset consisted of 750 CXR images (375 positive and 375 negative), whereas the validation dataset consisted of 135 CXR images (45 positive and 90 negative). The cases in the training and validation datasets were randomly selected. In contrast, our test dataset consisted of 402 CXR images (214 positive and 188 negative) with images from different manufacturers and with both fixed and portable characteristics. Therefore, our test dataset was suitable to perform sensitivity and specificity analysis for the different deep learning models. Of note, in our data split, we ensured that CXR images from the same patient stayed in the same partition (training, validation, and testing datasets) to avoid any biases.

\subsubsection{Deep Learning Methods}
We utilized four different state-of-the-art deep learning architectures (ResNet50, DenseNet161, InceptionV3, ResNeXt101) for the detection of pneumoperitoneum on CXR images \cite{he2016deep, iandola2014densenet, chen2017dual, szegedy2016rethinking}. We used pre-trained models on the ImageNet dataset \cite{deng2009imagenet} for each architecture to benefit from transfer learning in our training process. Utilizing transfer learning is critical for the optimization of deep learning models on a limited number of images, such as in our training dataset. In our training, we did not freeze any of the convolutional layers to fine-tune the CNN weights for extraction of pneumoperitoneum-related features. 

The CXR images are resized according to the required input size of different deep learning models, i.e., 299x299 pixels for InceptionV3 and 224x224 pixels for the rest of the models. All deep learning models were trained on a PyTorch framework \cite{paszke2019pytorch} using an NVIDIA Quadro RTX graphics processing unit with 48 GB memory. We experimented with different hyper-parameters such as learning rate, number of epochs, and data augmentation options for each model to minimize both training and validation losses. For the final models, we spent 100 epochs for training, which we found sufficient for the convergence of our optimization process on the dataset. We also tried different learning rates (1e-2 to 5e-4) for training the models in our study. Data-augmentation (horizontal flip, vertical flip, and random rotations from -15$^{\circ}$ to 15$^{\circ}$) was performed on the fly during training. In this training, we used binary cross-entropy as the loss function, a stochastic gradient descent optimizer, a batch size of 256, and a momentum value of 0.9. We reduced the learning rate by a factor of 0.1 after every 25 epochs. The final model was selected based on minimum validation loss during training.

\begin{table*}[]
	\centering
	\addtolength{\tabcolsep}{2pt}
	\renewcommand{\arraystretch}{1.5}
	\caption{The performance measures of various deep learning models for binary classification of pneumoperitoneum.}
	\label{result2s}
	\scalebox{0.85}
	{
		\begin{tabular}{lllllll} \hline
			Method      & Sensitivity & Specificity & Accuracy & Precision & F-1 Score & AUC   \\\hline\hline
			InceptionV3 & 0.841       & 0.931       & 0.883    & 0.932     & 0.884     & 0.938 \\
			ResNet101   & 0.873       & 0.936       & 0.902    & 0.937     & 0.906     & 0.946 \\
			ResNeXt101  & 0.865       & 0.952       & 0.905    & 0.953     & 0.907     & 0.951 \\
			DenseNet161 & 0.916       & 0.899       & 0.908    & 0.911     & 0.913     & 0.957\\ \hline
	\end{tabular}}
\end{table*}

\begin{figure}
	\centering
	\includegraphics[scale=.5]{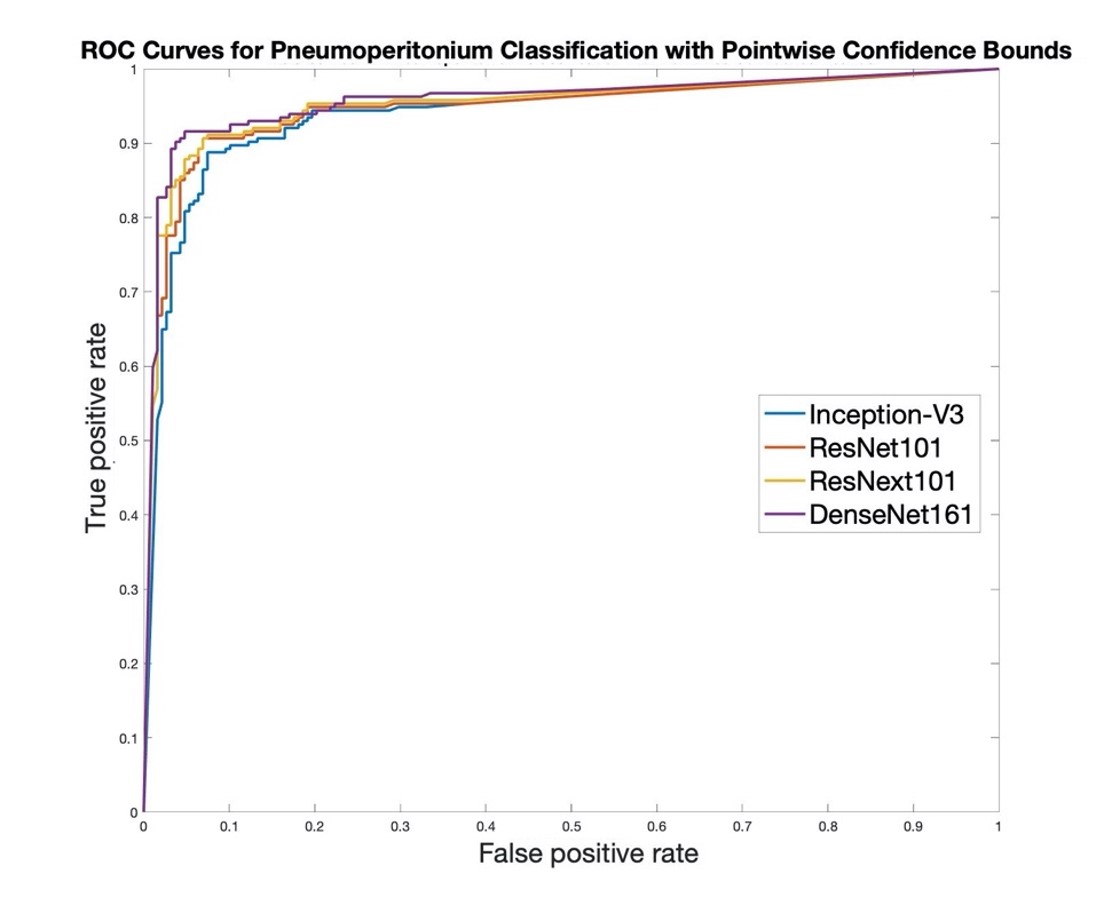}
	\caption{ROC curve for all deep learning models trained on the test dataset.}
	\label{fig:ROCM}
\end{figure}

\section{Results}
We evaluated the different deep learning models for the binary classification of pneumoperitoneum using our test dataset. In Table \ref{result2s}, we report sensitivity, specificity, accuracy, F1-score, and area under the receiver operating characteristic curve (AUC) as our evaluation metrics. These metrics are considered reliable measures for assessing the quality of machine learning models.
All deep learning models, particularly DenseNet161 and ResNeXt101, performed well for the binary classification of pneumoperitoneum. DenseNet161 achieved the highest accuracy of 0.908, whereas ResNeXt101 (0.905), ResNet101 (0.902), and InceptionV3 (0.883) performed slightly worse. For sensitivity, DenseNet161 (0.916) again outperformed ResNet101, InceptionV3, and ResNeXt101 by a margin of 0.43, 0.75, and 0.51, respectively. On the contrary, DenseNet161 achieved the lowest score of 0.899 for specificity, whereas ResNeXt101 performed best in this category, with a score of 0.952. The AUC score is considered to be a stable performance metric for evaluating machine learning approaches. ResNeXt101 and DenseNet161 achieved 0.951 and 0.957, respectively, for AUC. The ROC curves for all deep learning models are shown in Fig. \ref{fig:ROCM}.

\begin{table*}[]
	\centering
	\addtolength{\tabcolsep}{2pt}
	\renewcommand{\arraystretch}{1.5}
	\caption{Stratification of the Test Dataset (402 CXR images: 214 positive and 188 negative) according to the CXR image characteristics. Where CXRs is Chest X-rays, Pos is Positive cases, Neg is Negative cases, andd MFR. is Manufacturer.}
	\label{strat}
	\scalebox{1}
	{
		\begin{tabular}{lll} \hline
			Characteristic - Type         & Portable           & Fixed            \\\hline
			No. of CXRs                   & 318 CXRs           & 84 CXRs          \\
			Pos/Neg Cases                 & Pos - 183 \& Neg - 135 & Pos -31 \& Neg - 53 \\\hline
			Characteristic – MFR. & Philips            & Others           \\\hline
			No. of CXRs                   & 260 CXRs           & 142 CXRs         \\
			Pos/Neg Cases                 & Pos - 156 \& Neg - 135 & Pos - 58\& Neg - 84 \\ \hline
	\end{tabular}}
\end{table*}

\begin{table*}[]
	\centering
	\addtolength{\tabcolsep}{2pt}
	\renewcommand{\arraystretch}{1.5}
	\caption{The performance metrics for DenseNet161 and ResNeXt101 on the stratified test dataset according to portable/ fixed and Philips/Others Manufacturer characteristics.}
	\label{resultst}
	\scalebox{0.75}
	{
		\begin{tabular}{llllllll} \hline
			Dataset                   & Method      & Sensitivity & Specificity & Accuracy & Precision & F-1 Score & AUC   \\\hline\hline
			\multirow{2}{*}{Portable} & ResNeXt101  & 0.863       & 0.948       & 0.899    & 0.958     & 0.908     & 0.950 \\
			& DenseNet161 & 0.907       & 0.903       & 0.905    & 0.927     & 0.917     & 0.956 \\\hline
			\multirow{2}{*}{Fixed}    & ResNeXt101  & 0.871       & 0.962       & 0.929    & 0.931     & 0.900     & 0.974 \\
			& DenseNet161 & 0.968       & 0.887       & 0.917    & 0.833     & 0.896     & 0.958 \\\hline
			\multirow{2}{*}{Philips}                   & ResNeXt101  & 0.865       & 0.952       & 0.900    & 0.964     & 0.912     & 0.938 \\
			& DenseNet161 & 0.917       & 0.885       & 0.904    & 0.923     & 0.920     & 0.946 \\\hline
			\multirow{2}{*}{Others}                     & ResNeXt101  & 0.862       & 0.952       & 0.915    & 0.926     & 0.893     & 0.951 \\
			& DenseNet161 & 0.914       & 0.917       & 0.915    & 0.883     & 0.898     & 0.957\\\hline
	\end{tabular}}
\end{table*}

\begin{figure}
	\centering
	\includegraphics[scale=.52]{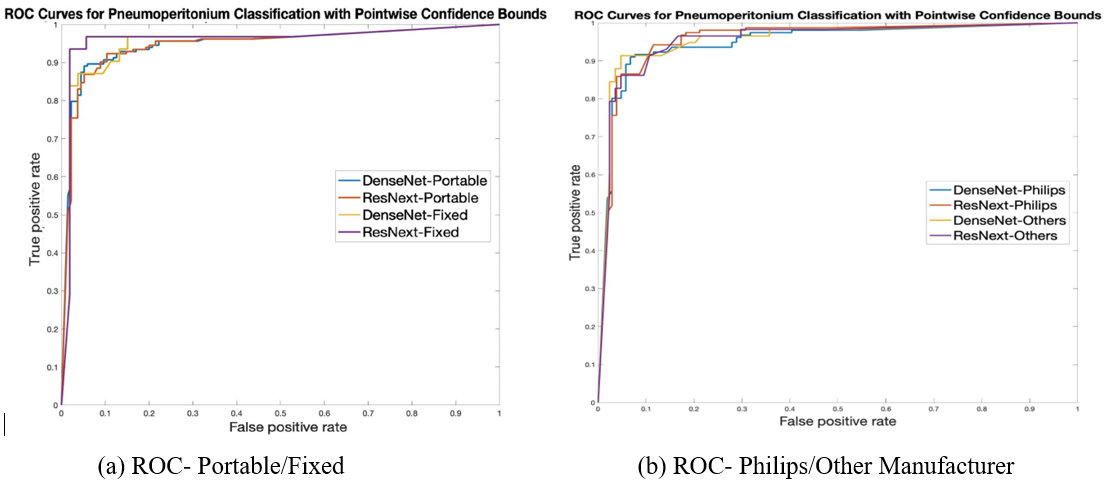}
	\caption{ROC curves for DenseNet161 and ResNeXt101 for the stratified test dataset based on (a) Portable/Fixed and (b) Philips/Other Manufacturer.}
	\label{fig:ROCS}
\end{figure}

\subsection{Stratified Test Results}
As shown in Table \ref{strat}, we stratified the test dataset by image characteristics, such as different manufacturers and fixed/portable imaging systems, to further evaluate the effect of these characteristics on our best performing deep learning models (DenseNet161 and ResNeXt101). We found that the performance of both of these models on a stratified test dataset was comparable to the full test dataset, as shown in Table \ref{resultst}. However, there were a few exceptions. For instance, DenseNet161 achieved a higher sensitivity of 0.968, and ResNeXt161 also showed a higher AUC score of 0.974 on images in the fixed imaging system subgroup of the test dataset. This is likely due to a small sample size of 84 CXR images in this subset. The ROC curves for these stratified test datasets are shown in Fig. \ref{fig:ROCS}.

\begin{figure}
	\centering
	\includegraphics[scale=0.55]{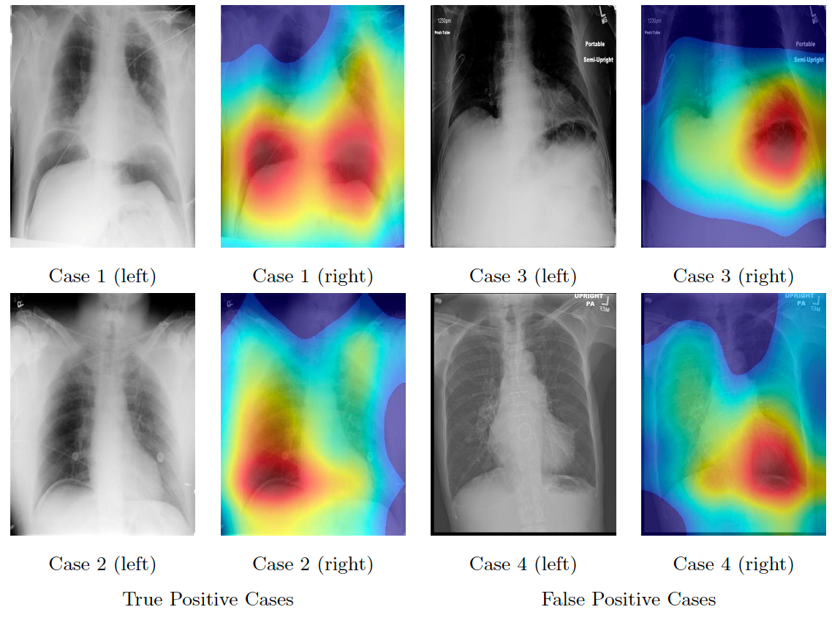}
	\caption{Examples of Grad-CAM Activation of true-positive (1 and 2) and false-positive (3 and 4) cases by DenseNet161: the left image of each case is the original image, whereas the right image is an activation by the Grad-CAM method. The red coloring indicates a highly weighted region of interest. In the top true positive case (1), the pneumoperitoneum (free air) is beneath both the right and left diaphragm, and the heat map correctly marks both sides. In the bottom case (2), free air is only under the right diaphragm, and the heat map identifies it on the correct side. The false positive cases (3 and 4) in these examples are due to air in the bowel, below the left diaphragm (in the upper images), and air in the stomach, below the left diaphragm (in the lower images), without pneumoperitoneum in either case.}
	\label{fig:Vis}
\end{figure}

\subsection{Model Visualization and Error Analysis}
We used the Grad-CAM algorithm \cite{selvaraju2017grad}, which uses pneumoperitoneum specific gradient information flowing into the final convolutional layer of the DenseNet161 deep learning model to mark the regions of interest on the CXR images that heavily influenced the outcomes of our model. Examples of Grad-CAM activations on randomly selected true positive cases of pneumoperitoneum are shown in Fig. \ref{fig:Vis}. This visualization produces localization maps of the regions of interest for CXR images and can provide an explanation for the final diagnostic decisions of the deep learning models. The red coloring indicates the most important regions for the ultimate decision of the model on CXR images. We also applied the Grad-CAM algorithm on randomly selected false-positive cases of pneumoperitoneum, which were incorrectly identified by our DenseNet161 model. We found that false-positive cases, CXRs without pneumoperitoneum, were most frequently due to air in the stomach or small bowel below the left diaphragm, or lucency in the lungs above the diaphragm, as shown in Fig. \ref{fig:Vis}.

\section{Discussion}

In this study, we used various common deep learning architectures to develop a model for binary classification of pneumoperitoneum on chest X-ray exams. Pneumoperitoneum, also known as free air, is the abnormal presence of air in the peritoneal cavity. In this experiment, we developed our deep learning models using training and validation datasets that only consisted of CXR images from Philips fixed imaging systems, whereas, in the test dataset, we used CXR images from portable imaging systems or from other imaging system manufacturers (Siemens, Kodak, etc.). Our experiment showed that deep learning models trained on data from a fixed imaging system from a single institution performed well on heterogeneous data from other institutions. Particularly, our deep learning models in this study achieved a high specificity and sensitivity on our diverse test dataset overall. 

In our stratified test dataset, the two best performing models (DenseNet161 and ResNeXt101) showed consistent performance regardless of various image characteristics, such as fixed, portable, Philips, or other manufacturer. ResNet is a residual network that uses skip or shortcut connections to allow one to pass the information over convolutional blocks to extract abstract features. In contrast, DenseNet is a densely connected convolutional network that simplifies and refines the connection between convolutional blocks to ensure the flow of maximum information and gradients through feature reuse, i.e., features extracted by very early layers are directly used by deeper layers.

The models showed slightly higher performance in the fixed imaging system subgroup. This may be attributed to a small number of CXR images (84) in this subgroup. Furthermore, the Grad-CAM algorithm showed that our models accurately identify the correct anatomic area and features on the CXR images for the detection of pneumoperitoneum. Radiologists can generally identify and interpret the urgency of the findings based on chart review of patients' medical history. However, the goal of this deep learning approach for pneumoperitoneum detection is to identify and triage possible urgent cases for interpretation rather than replacing the need for radiologist interpretation. A major application for our deep learning algorithm is to screen and triage all imaging with critical findings for expedited interpretation and patient care. Particularly, when the reading list is long, such deep learning approaches can assist with prioritizing urgent exams, especially when the finding is not tagged as STAT (immediate) priority. In addition, when there are many other findings in chest X-rays, subtle pneumoperitoneum cases can be missed. The proposed deep learning model can help radiologists by drawing attention to those cases. 

\paragraph{This study has several limitations} First, our study would benefit from further validation on a larger external test dataset and a prospective clinical trial, which we will pursue as future work. Second, our pipeline is focused on distinguishing pneumoperitoneum negative and positive cases, and does not recognize other urgent or critical findings such as pneumothorax or pneumonia on chest X-rays, and such findings could require patients to seek immediate medical or surgical attention. As future work, we plan to include other urgent findings in the next version of our model and will evaluate it in a multi-class classification setting. Finally, although we used Grad-CAM in this study to visualize the regions of interest in our classification, we plan to develop and evaluate a precise detection and a segmentation model to localize pneumoperitoneum on CXR images.

\section{Conclusion}
In summary, this study evaluated the generalizability of deep learning models across different image characteristics for the detection of pneumoperitoneum on CXR images. Our results showed that end-to-end deep learning models performed well in detecting pneumoperitoneum on CXR images from different types of imaging systems at various institutions. If clinically validated, this system could assist radiologists as a pre-screening tool to help prioritize chest X-rays with emergent findings, or offer a second opinion for the presence of pneumoperitoneum on CXR images. For future study, we plan to expand our training dataset to a large multi-institutional dataset to further improve the performance of various deep learning models selected for this task. Also, we plan to expand our test dataset and run prospective clinical trials for further validation of our models. Finally, we plan to expand our study to include other imaging modalities, such as CT scans, to assist with the detection of other urgent and critical findings detected on radiology exams. 

\section*{Ethical Considerations}
The use of human subject data in this study was approved by the Dartmouth Institutional Review Board (IRB) with a waiver of informed consent.

\section*{Conflicts of interest}
None Declared

\section*{Funding}
This research was supported in part by National Institute of Health grants R01LM012837 and R01CA249758.

\bibliography{mybibfile}

\end{document}